\documentclass[parallelisme]{compas2023}

\toappear{1} % Conserver cette ligne pour la version finale

\usepackage{url, color}
\usepackage{caption}
\usepackage{subcaption}

\usepackage{xcolor}

\begin{document}

\title{Towards Sustainable Low Carbon Emission Mini Data Centres}
\shorttitle{}

\author{Ismael Samaye, Paul Leloup, Gilles Sassatelli, Abdoulaye Gamati{\'e}}%

\address{LIRMM, Université de Montpellier - CNRS\\ 
Montpellier, France\\
first.last@lirmm.fr}

\date{\today}

\maketitle

%===========================================================         %
%R\'esum\'e
%===========================================================  
\begin{abstract}
Mini data centres have become increasingly prevalent in diverse organizations in recent years. They can be easily deployed at large scale, with high resilience. They are also cost-effective and provide high-security protection. On the other hand, IT technologies have resulted in the development of ever more energy-efficient servers, leading to the periodic replacement of older-generation servers in mini data centres. However, the disposal of older servers has resulted in electronic waste that further aggravates the already critical e-waste problem. Furthermore, despite the shift towards more energy-efficient servers, many mini data centres still rely heavily on high-carbon energy sources. This contributes to data centres' overall carbon footprint. All these issues are concerns for sustainability. 
In order to address this sustainability issue, this paper proposes an approach to extend the lifespan of older-generation servers in mini data centres. This is made possible thanks to a novel solar-powered computing technology, named Genesis, that compensates for the energy overhead generated by older servers. As a result, electronic waste can be reduced while improving system sustainability by reusing functional server hardware. Moreover, Genesis does not require server cooling, which reduces energy and water require\-ments. Analytical reasoning is applied to compare the efficiency of typical conventional mini data centre designs against alternative Genesis-based designs, in terms of energy, carbon emissions and exploitation costs.

  \MotsCles{Sustainability, mini data centre, renewable energy, server refresh, server consolidation, solar energy}
\end{abstract}

\section{Introduction}
Recent advances have been made in information technology, particularly in data centres, which provide cloud services, e.g. the Internet of Things, streaming, etc. 
However, this sustained progress comes at a price, especially for the environment, including an increase in carbon dioxide ($CO_2$) emissions as a result of the production of electricity to answer the energy demand for data centres, which accounts for approximately 1\% of worldwide global energy consumption, as indicated in \cite{Art:1}. Data centres commonly use non-green energy sources that contribute to significant $CO_2$ emissions \cite{Art:2}, which account for $0.3\%$ of the total $CO_2$ emissions on the planet. Besides $CO_2$ emissions, data centres are also concerned about water stress problems in some regions \cite{Art:3}.

\begin{figure}[htbp]
  \centering
  \includegraphics[width=10cm,height=8cm]{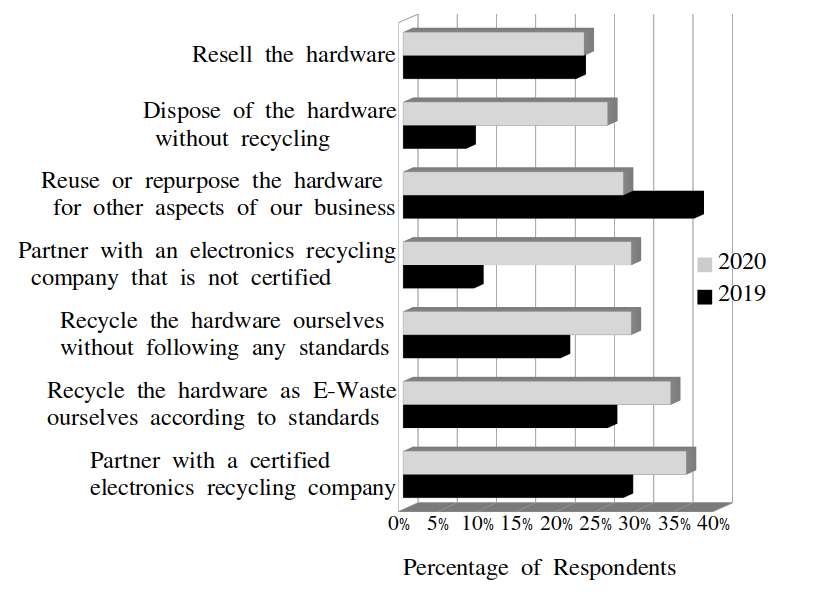}
  \caption{\small Survey of e-waste plans in 2019 and 2020  \cite{supermicro:2018}}
  \label{fig:ewaste}
\end{figure}

\vspace{.1cm}
\textbf{On the environmental impact of data centre equipment.}
Gupta \textit{et al.}  pointed out that data centre hardware and infrastructure also significantly contribute to the amount of $CO_2$ released \cite{Art:4}. Electronic waste is another pollution source. Indeed, hazardous substances may be present in metals and electronic components retrieved from data center equipment and deposited in the environment. Within this category, there are two distinct types of equipment: data center infrastructure that comprises mechanical and electrical components, including transformers, generators, air conditioners, racks, and cables; and computing components, such as servers. Equipment is periodically replaced to improve performance and energy efficiency \cite{Art:5}. Infrastructure components are replaced every 10 to 15 years. In contrast, computing components are more frequently changed. A white paper from the SuperMicro Company \cite{supermicro:2018} examines the server renewal cycle in data centres. Electronic waste management approaches are also discussed. The paper argues that green IT is driven by the replacement of older systems with newer ones that can handle more intensive workloads. 
In 2020, only $68\%$ of surveyed respondents renewed their servers annually or every 2 to 3 years. This is compared to only $37\%$ in the 2019 survey. For those who keep their servers for six years or more, only $8\%$ reported doing so in 2020, compared to $23\%$ in the 2019 survey. For organizations that refresh very frequently, every year, $7\%$ did so in the 2019 survey. However, this number increased significantly to $26\%$ in the 2020 survey. This enthusiasm for new, more energy- and workload-efficient systems is a trap. Anticipating the widespread adoption of such a system, equipment replacement frequency is likely to surge. This will increase the volume of e-waste and exacerbate the environmental impact of mineral extraction and server manufacturing. 
Regarding the electronic waste issue, there are various approaches surveyed by \cite{supermicro:2018}, which are shown in Figure \ref{fig:ewaste}. One can observe an increase in efforts to mitigate the e-waste issue in 2020 compared to 2019.

\vspace{-.6cm}
\vspace{-.6cm}

\vspace{.1cm}
\textbf{Problem statement and our contribution.}
Reusing servers in data centres has become a critical issue for modern businesses and organizations. It contributes to cost reduction and data centre sustainability. Additionally, it can help reduce environmental impact by reducing the need for new servers and limiting waste. However, server reuse should not compromise data centre security (such as hosted data confiden\-tiality) or quality. Therefore, a careful balance should be found between security and environmental impacts due to server reuse. The aforementioned issues have been mostly studied on large data centres, but not on mini data centres which are widespread nowadays. Approximately 40\% of all data centres in the United States are mini data centres \cite{Mini:Mohan}. Generally, a mini data centre consists of less than 1,000 square feet of space in which servers are located. It typically has fewer than twenty-five servers \cite{Mini:Mohan}. 
The server utilization rate of conventional data centres usually ranges from 6-12\% \cite{time}. A major benefit of mini data centres is scalability: they can be deployed more easily and can be scaled up or down as per the organization's needs. Moreover, mini data centres can be distributed across different locations, making them more resilient to disruptions such as natural disasters or power outages. This enforces resource availability. Finally, they can be designed with security in mind, e.g., easy isolation from other IT infrastructures. 

\quad The sustainability issue addressed in the present work concerns server lifetime extension in mini data centres. This is done while mitigating the associated energy efficiency penalty from server aging. Indeed, older servers tend to be less energy-efficient than newer ones. In order to resolve this issue, we adopt a novel computing technology, called Genesis \cite{Genesis:Prof, GAMATIE2023100844}, which enables us to deploy mini data centres powered by solar energy in a similar way to smart grids. Using this free energy under favorable meteorological conditions can compensate for energy consumption due to old-generation servers kept in the system. Therefore, Genesis could help solve the problem of renewing servers, which generate electronic waste. Furthermore, it reduces data centre operating costs, e.g., electricity bills, with low-carbon emissions. We use an analytical method to show the potential benefits of Genesis against conven\-tional mini data centre designs w.r.t. server renewal.

\section{Related work}
Many studies have been conducted on server renewal and lifespan extension. Consolidation of a limited number of powerful servers is an approach often adopted for reducing physical machines in data centres. It can reduce operational expenses. Qiu \textit{et al.} proposed using a probabilistic request allocation problem to minimize the number of servers required to meet workload execution requirements without violating service level agreements \cite{Qiu}. Farahnakian \textit{et al.} studied a virtual machine (VM) consolidation approach by considering a regression-based model that approximates the future CPU and memory utilization of VMs \cite{7593250}.

\quad Doyle \textit{et al.} \cite{joseph} addressed the issue of finding the most appropriate periods for server renewal in mini data centres, so as to maximize the return on investment. They presented a comprehensive framework allowing to evaluate selected server deployment scenarios according to various hardware refresh options. They studied several case studies of mini data centre configurations where companies achieved a \textit{return on investment} (ROI) by upgrading their computing hardware at the right moment. Several key factors to consider when upgrading are discussed. This upgrade is also discussed in terms of its benefits for small data centres. 
They confirmed the benefits of consolidation, i.e., aggregation of multiple compute or networking services onto a lesser number of hardware nodes.

\quad In \cite{Bash}, Bashroush also showed that optimizing the computing infrastructure through hardware refresh or increasing its utilization can result in significant energy savings in a data centre than reducing the power usage efficiency (PUE). He proposed equations and models to guide refresh strategies to reduce environmental impact and operating costs. In \cite{Art:5}, Bashroush \textit{et al.} rather focused on electronic waste using a different server refresh strategy. Instead of replacing the whole servers for upgrading the compute infrastructure, they advocate renewing only some components within servers, e.g., faulty disks as reported in a Google data centre \cite{ssd}, thus improving the performance of the servers while decreasing electronic waste. They upgrade machines in use for five to six years. This extends servers' lifespan and reduces electronic waste dumped into nature. To better understand the server failure issue, a study has been carried out on over 290,000 hardware component failures recorded over the past four years in \cite{Art}. Data is collected from hundreds of thousands of servers in dozens of data centres. Statistical analysis is applied to the collected data set to uncover failure characteristics across temporal, spatial, product range and component dimensions. By doing so, data centres can support partial component replacement. While such an approach is beneficial, some electronic waste remains. 

\quad Wang \textit{et al.} \cite{Sustainability-aware:2020} dealt with the reduction of $CO_2$ emissions by suggesting early actions during data centre building beyond the operational phase. So, they propose a novel manufacturing cost-aware server provisioning plan to mitigate the carbon footprint of the data centre construction phase. 

Vasconcelos \textit{et al}. \cite{silvavasconcelos:hal-04032094} studied the size of renewable energy sources for geographically distributed data centres around the world. They take into account the carbon footprint caused by electricity consumption in regions where data centres are deployed. In addition, they consider the carbon footprint associated with solar panels and battery manufacturing. They do not address the environmental impacts caused by server renewal in these distributed data centres.

\quad The aforementioned approaches are intended to lower data centres' energy consumption or electronic waste generation. However, they do not enable the aggressive reuse of old servers at a cost-effective energy consumption level for mini data centres. With Genesis, this is possible while using renewable energy sources for a lower carbon footprint.

\section{The Genesis technology}
\begin{figure*}[htbp]
     \centering
     \hspace{-1.5cm} 
     \begin{subfigure}[b]{0.6\textwidth}
         \centering
        \includegraphics[width=7.5cm,height=5.2cm]{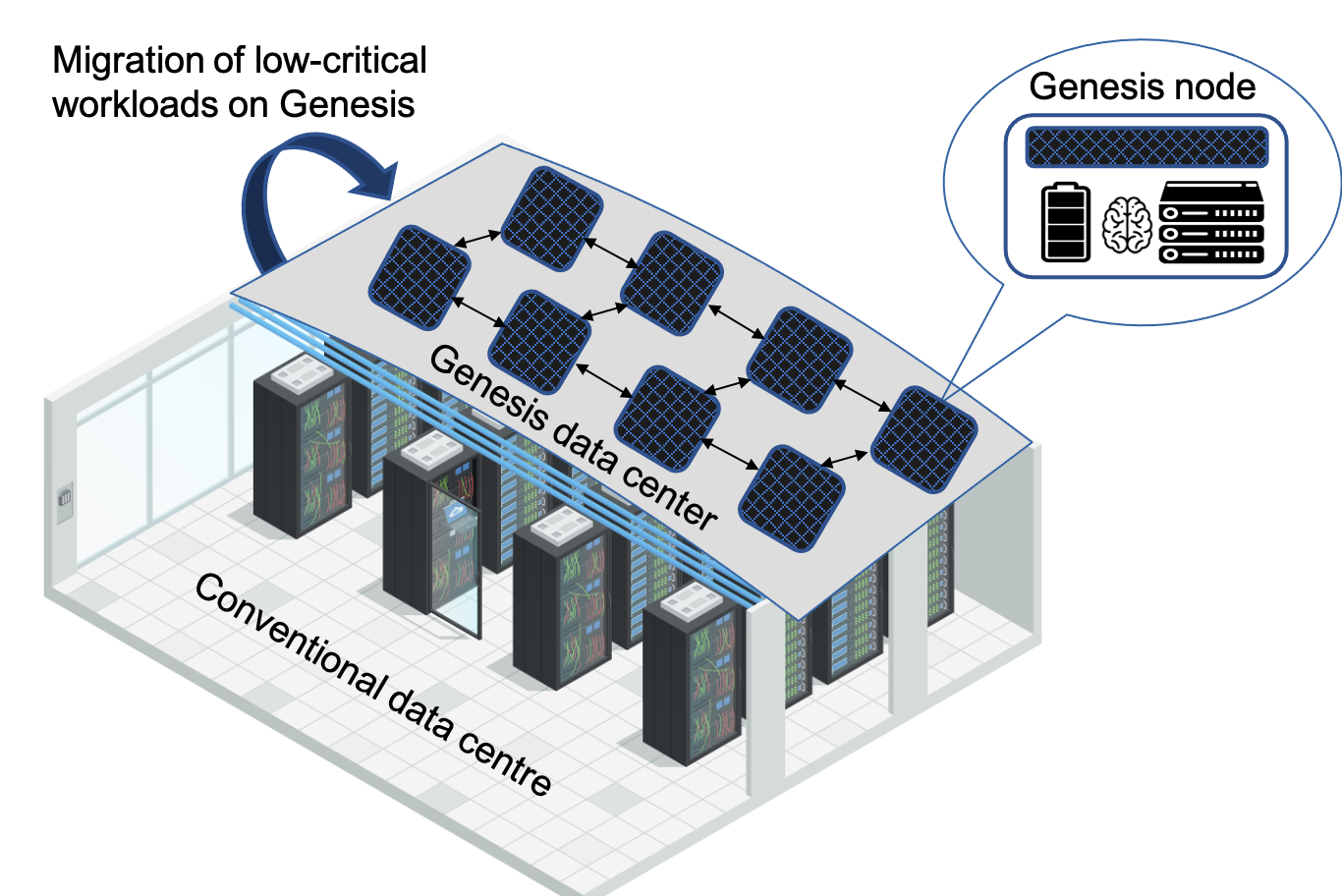}
         \caption{General view of Genesis system deployment}
         \label{fig:genesis:focus}
     \end{subfigure}
     \hspace{-0.5cm}
     \begin{subfigure}[b]{0.35\textwidth}
         \centering
            \includegraphics[height=4cm]{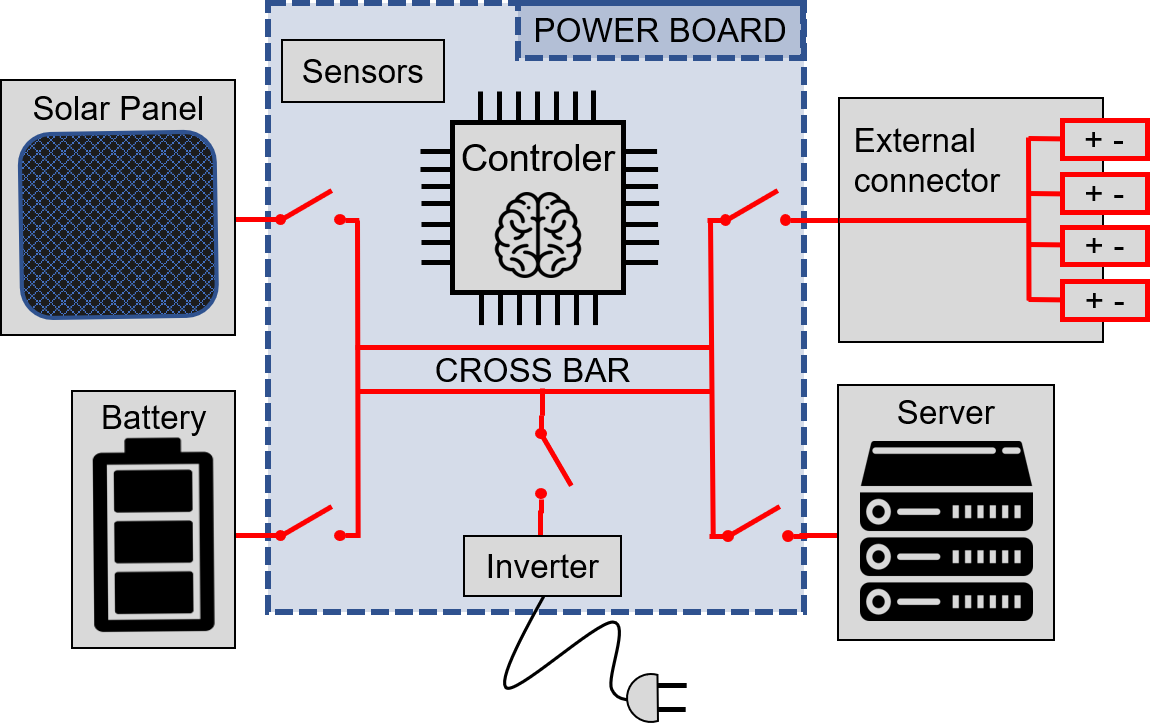}
         \caption{Zoom in a Genesis node}
         \label{fig:node}
     \end{subfigure}
        \caption{Integration of Genesis with conventional data centre for sustainable compute infrastructure (illustration partly based on macrovector/Freepik)}
        \label{fig:genesis}
\end{figure*}

We consider a novel computing technology, Genesis, recently implemented through a solar-powered mini data centre demonstrator \cite{Genesis:Prof}. As illustrated in Figure \ref{fig:genesis}, Genesis can be seen as a companion mini data centre typically deployed on the roof of a classical data centre building, in order to harvest the solar energy directly used in the system. Genesis is composed of modules or nodes that are interconnected. Each module can host computing resources and energy harvesting and storage facilities. Typically, it includes a solar panel, a battery, a server, and local power control logic for inter-node energy exchange. 
The modules can exchange energy and computational data, e.g., virtual machines. When a module lacks energy in its local battery to execute a workload, it can request energy from a remote module. Alternatively, it can send its local workload to another module having sufficient energy and compute capability for remote execution.  

\quad In the Genesis deployment depicted in Figure \ref{fig:genesis}, non-urgent or non-critical workloads, e.g., workloads with low quality of service (QoS) requirements, can be migrated from the indoor classical mini data centre to the on-roof Genesis data centre. The migrated workloads would therefore run on older servers repurposed by Genesis and powered by free solar energy. This integration between Genesis and a conventional mini data centre establishes a robust configuration that harnesses renewable energy for computing, except when the batteries are depleted. Note that Genesis could be deployed alone, even without access to the power grid (i.e. extreme conditions), operating only on harvested solar energy \cite{DBLP:conf/date/GamatieSM21}.

\vspace{-.6cm}
\vspace{-.6cm}

\quad Using our lab prototype of Genesis, we evaluated the integration of two typical old-generation servers into modules: a server with an Intel Xeon processor E5-1620 v2 with 10M cache and operating at 3.70 GHz, and a Dell PowerEdge T20 server. The former server was used six years ago in the world-class supercomputer hosted by CINES in Montpellier, France (\url{https://www.cines.fr/en}). The latter belongs to an old Dell server generation. Genesis' benchmarked configuration comprises a 2 kWh battery and 300 W solar panel per module. Given these configurations, a node integrating the most powerful server can run for 10h and 6.5h at 50\% and 100\% CPU stress levels respectively. This is achieved by only using the associated battery budget. When considering the less powerful Dell server option, a node can run for 27h and 19h at 50\% and 100\% stress levels respectively. Note that under favorable solar irradiation conditions, an empty node battery can be completely filled with energy within about 18 h, using a single solar panel. 

\quad Genesis can therefore offer a credible alternative to typical data centre workloads while favoring better sustainability. In the next sections, we rely on some extrapolated deployments of Genesis to analyze the possible benefits of this technology w.r.t. server renewal and reuse, in order to answer the data centre sustainability issue. These deployment scenarios assume that server reuse does not compromise data quality or security.

\section{Evaluation of different strategies for sustainable data centres}

We present a number of strategies for enabling sustainable mini data centres, under performance and energy-efficiency constraints. We start with a given mini data centre design corresponding to a typical system deployment scenario, $C_0$. Then, we explore various evolutions of $C_0$ over time, by applying state-of-the-art solutions, including the Genesis-based approach. Finally, we assess the efficiency of each scenario with respect to energy and carbon dioxide emissions. For this purpose, we consider the metrics defined below.

\quad Given $N$ servers in a mini data centre $DC$, we define the global\footnote{For simplicity, this is a rough definition, which does not include the energy consumption of all the other equipment of a data centre.} energy consumed by $DC$ as follows: 
\begin{equation} 
\label{eq1}
    Energy^{DC} = %(N * P * T ) + E_r 
    \sum_{i=1}^{N} (P_i * T) + E_{cooling}
    \end{equation}
where $P_i$ is the thermal design power (TDP) of each server $i$, T is the activity period of the servers during which some workloads are executed, and $E_{cooling}$ is the energy required for server cooling in the data centre.

\quad On the other hand, we define the global computational capacity of a data centre composed of $N$ servers as follows: 
\begin{equation} 
\label{eq2}
    Compute\_capacity^{DC} = 
    %N * P_s
    \sum_{i=1}^{N} Perf_i 
    \end{equation}
where $Perf_i$ represents the performance of a server $i$, expressed in FLOPs. 

For the sake of simplicity, the analytical reasoning we apply in the next sections makes some strong assumptions: i) the server operation cost is reduced to that of its integrated processors, i.e., other relevant components such as memory, storage, or GPUs are deliberately neglected here; ii) regarding server operations, we assume an on-off execution modality, i.e., when a server executes a workload it operates at 100\% of its maximum TDP, otherwise it consumes 0 W.

\subsection{Hardware renewal approaches based on conventional server technologies}

Table \ref{tab:perform} summarizes different characteristics of Intel server processors, for different processor generations \cite{joseph}. We can observe improvements in successive processors over time. This hardware evolution calls for server renewal in data centres for better energy efficiency and management costs. In the same table, we added the respective prices of the processors according to the online information found on the Intel website. 
Table \ref{tab:params} summarizes the configuration parameters of the analyzed mini data centre designs. We highlight the number and types of servers, as well as the solar panels and batteries for each configuration.

\begin{table}[htbp]
    \centering
    \footnotesize
   \begin{tabular}{|p{2cm}|p{1.4cm}|p{1.3cm}|p{2.2cm}|p{2.2cm}||p{1.7cm}|}
\hline
Year of Release & Processor type & TDP (W) & Performance per server (GFLOPs) & Energy efficiency (GFLOPs / watt) & Online prices from Intel (\$)  \\
\hline
2011 & i7-2600 & 95  & 32.52& 0.3423 & 135.40\\
2012 & i7-3770 & 77  & 29.29 & 0.3804 & 189.15\\
2013 & i7-4770 & 84  & 31.57 & 0.3758 & 137.74\\
2015 & i7-6700 & 65  & 31.42 & 0.48334 & 311.86\\
2017 & i7-7700 & 65  & 34.14 & 0.5252 & 343.33\\
2018 & i7-8700 & 65  & 92.07 & 1.4165 & 287.80\\
\hline
\end{tabular} 
    \caption{Performance of Intel Core i7 Processors \cite{joseph} and server acquisition cost}
    \label{tab:perform}
\end{table}

\begin{table}[h]
\footnotesize
    \centering
   \begin{tabular}{|p{1.8cm}||p{1.3cm}|p{3.0cm}|p{1.8cm}|p{2.8cm}|p{1.6cm}|}
\hline
 System config.& \# Servers & Processor type &  \# Solar panels & \# Batteries & Power (kW) \\
\hline
$C_{ref}$ & 12 & i7-2600 & 0 & 0 & 1.14\\
\hline
$C_{0}$ & 12 & i7-4770 & 0 & 0 & 1.01\\
\hline
$C_{1}$ & 5 & i7-8700 & 0 & 0 & 0.33 \\
\hline
$C_{2}$ & 5 & i7-8700 & 5 & 2kWh $\times 5$ & 0.42\\
\hline
$C_{3}$ & 14 & i7-8700 $\times 5$  \&  i7-4770 $\times 9$ & 14 & 2kWh $\times 14$ & 1.36\\
\hline
$C_{4}$ & 7 & i7-8700 $\times 5$  \&  i7-4770 $\times 2$ & 7 & 1kWh $\times 5$ \& 2kWh $\times 2$ & 0.63 \\
\hline
$C_{5}$ & 7 & i7-8700 $\times 4$ \&  i7-4770 $\times 3$ & 7 & 1kWh $\times 4$ \& 2kWh $\times 3$ & 0.65\\
\hline

\end{tabular} 
    \caption{Configuration parameters of mini data centre}
    \label{tab:params}   
\end{table}

\quad Let us consider a reference configuration $C_{ref}$ of mini data consisting of 12 old-generation i7-2600 servers. The corresponding energy consumption and computation capacity are given in Table \ref{tab:analyse1}. Here, since mini data centres are utilized between 40\% and 50\% \cite{joseph}, we assume a daily execution time of 10 hours. On the other hand, we consider the energy corresponding to data centre cooling to represent 7\% of servers' energy consumption \cite{joseph}. 

\begin{table}[h]
\footnotesize
    \centering
   \begin{tabular}{|p{2.cm}||p{2.5cm}|p{4.5cm}|}
\hline
 System config.& $Energy^{DC}$ (kWh) &$Compute\_capacity^{DC}$ (GFLOPs)\\
\hline
$C_{ref}$ & 12.19 & 390.24 \\
\hline
$C_0$  & 10.78 (-11.6 \%) & 378.84 (-2.92 \%)\\
\hline
$C_1$ & 3.47  (-71.55 \%) & 460.35  (+17.96 \%)\\
\hline
\end{tabular} 
    \caption{Reference configuration $C_{ref}$ versus improved configuration $C_0$ and $C_1$ }
    \label{tab:analyse1}   
\end{table}

\quad To improve the energy consumption and computation capacity of a mini data centre, the first solution consists in replacing each server with a more powerful server with a better TDP and possibly better performance per server. In the specific case of the above reference configuration, one can consider a data centre server renewal composed of twelve i7-4770 servers (configuration $C_0$). Then, the obtained energy gain and computation capacity loss compared to $C_{ref}$ are respectively 11.6\% and 2.92\% (see Table \ref{tab:analyse1}). 
Alternatively, one can apply a different server renewal approach consisting of consolidation \cite{joseph} by using only 5 i7-8700 servers for the entire mini data centre (configuration $C_1$). Now, the observed energy and computation capacity gains compared to $C_{ref}$ are respectively 71.55\% and 17.96\% (see Table \ref{tab:analyse1}). Overall, we observe that server consolidation as suggested in the configuration $C_1$ provides the highest outcome both in terms of energy and computation capacity. The same observation can be found in \cite{joseph}.

\quad While server consolidation offers interesting energy reductions while improving computation capacity, its implementation in practice requires suitable mechanisms to mitigate cooling requirements. Indeed, compared to a configuration such as $C_{ref}$, the consolidated configuration $C_1$ consists of fewer but more powerful servers that can result in increased heat dissipation. Cooling systems consume considerable energy. For instance, Doyle \textit{et al.} in \cite{joseph} assume that the maximum allowable temperature in a consolidated mini data centre based on the servers highlighted in Table \ref{tab:perform} is 45°C. To ensure this, they considered a mechanism consuming 7\% of server energy consumption, without accounting for fan power consumption. From a different perspective, such mechanisms induce additional investment costs, while the discarded old-generation servers may generate a crucial sustainability problem, especially when consolidation concerns a high number of servers in the reference mini data centre configuration.
 
\subsection{Hardware renewal approaches based on the Genesis technology}

Considering Genesis technology for data centre energy reduction and old server refresh issues, we explore previous hardware renewal issues. Given the specific characteristics of Genesis, the energy consumption definition of a data centre is slightly different from that of the formula (\ref{eq1}), as follows: 
\begin{equation} 
\label{eq3}
    Energy_{DC} = ((\sum_{i=1}^{N} P_i) + Overhead) * T
    \end{equation}
 where $N$ is the number of servers in the system, $P_i$ is the thermal design power of a server $i$, $T$ is the activity period of the servers, and $Overhead$ represents the static power of a Genesis infrastructure except the server. The outdoor Genesis system benefits from passive cooling, so there is no cooling energy needed.

\quad In the following, we analyze new mini data centre configurations based on Genesis w.r.t. energy consumption and computation capacity. This is done when integrating old and new-generation servers into Genesis modules. 
We assume each module can include a full 2 kWh or 1 kWh battery with usable energy of 1.7 kWh and 0.85 kWh per battery. The Genesis prototype's power overhead is around 20W. It corresponds to infrastructure power management logic.

\quad Table \ref{tab:analyse2} summarizes the outcome of the above configurations based on Genesis compared to the conventional reference configuration $C_{ref}$. In $C_2$, each battery in a module is charged with 2 kWh solar energy, where 85\% of the stored energy is exploitable for computing workloads. Given the five modules in this Genesis configuration, the initially available energy is 8.5 kWh. It largely covers the energy consumption of the considered 5 i7-8700 servers used for consolidation, i.e. 4.2 kWh (this energy includes the static power overhead of the infrastructure). 

\begin{table}[h]
\footnotesize
    \centering
   \begin{tabular}
{|p{2.2cm}|p{2.5cm}|p{4.5cm}|}
\hline
 System config.&  
 Energy$^{DC}$ (kWh) & $Compute\_capacity^{DC}$ (GFLOPs)\\
\hline
$C_{ref}$ & 
12.19 & 390.24 \\
\hline
$C_2$ & 
4.25 & 460.35 \\
\hline
$C_3$ & 
13.61 & 744.48\\
\hline
$C_4$ &
6.33 & 523.49 \\
\hline
$C_5$ & 
6.52 & 462.99\\
\hline
\end{tabular} 
    \caption{Genesis with new servers $C_2$ and Genesis with new and servers ($C_3$, $C_4$ and $C_5$)}
    \label{tab:analyse2}
\end{table}

\quad Given the substantial residual green energy in the batteries from $C_2$, there is the potential to use this energy for expanding the computation capacity of the mini data centre. The $C_3$ configuration is therefore extended to 14 modules, each potentially containing a battery. With this increase in data centre resources, not only the available green energy increases but also the computing capacity. Another side effect of the configuration $C_3$ is the possibility to continue reusing old-generation servers together with new-generation servers to extend the lifetime and reduce the environmental impact of older servers.

\quad A further resource optimization of the configuration $C_3$ can, however, be achieved. For instance, in configuration $C_4$ we assume five Genesis modules, each equipped with a 1 kWh battery. The usable energy per module is 0.750 kWh for i7-8700 servers and two modules with 2 kWh battery for older i7-4770 servers. The resulting configuration is well-sized, ensuring high computation capacity. Finally, in the last configuration $C_5$, we adjust $C_4$ by slightly reducing the number of i7-8700 servers and increasing the i7-4770 servers. The resulting computing capacity is still higher than in $C_{ref}$ while the associated energy consumption is covered by the green energy budget available in the different batteries. 

\section{Comparison of design efficiency: energy, carbon and cost}
\begin{figure*}[htbp]
     \centering
     \begin{subfigure}[htbp]{0.45\textwidth}
         \centering
           \includegraphics[height=4.5cm]{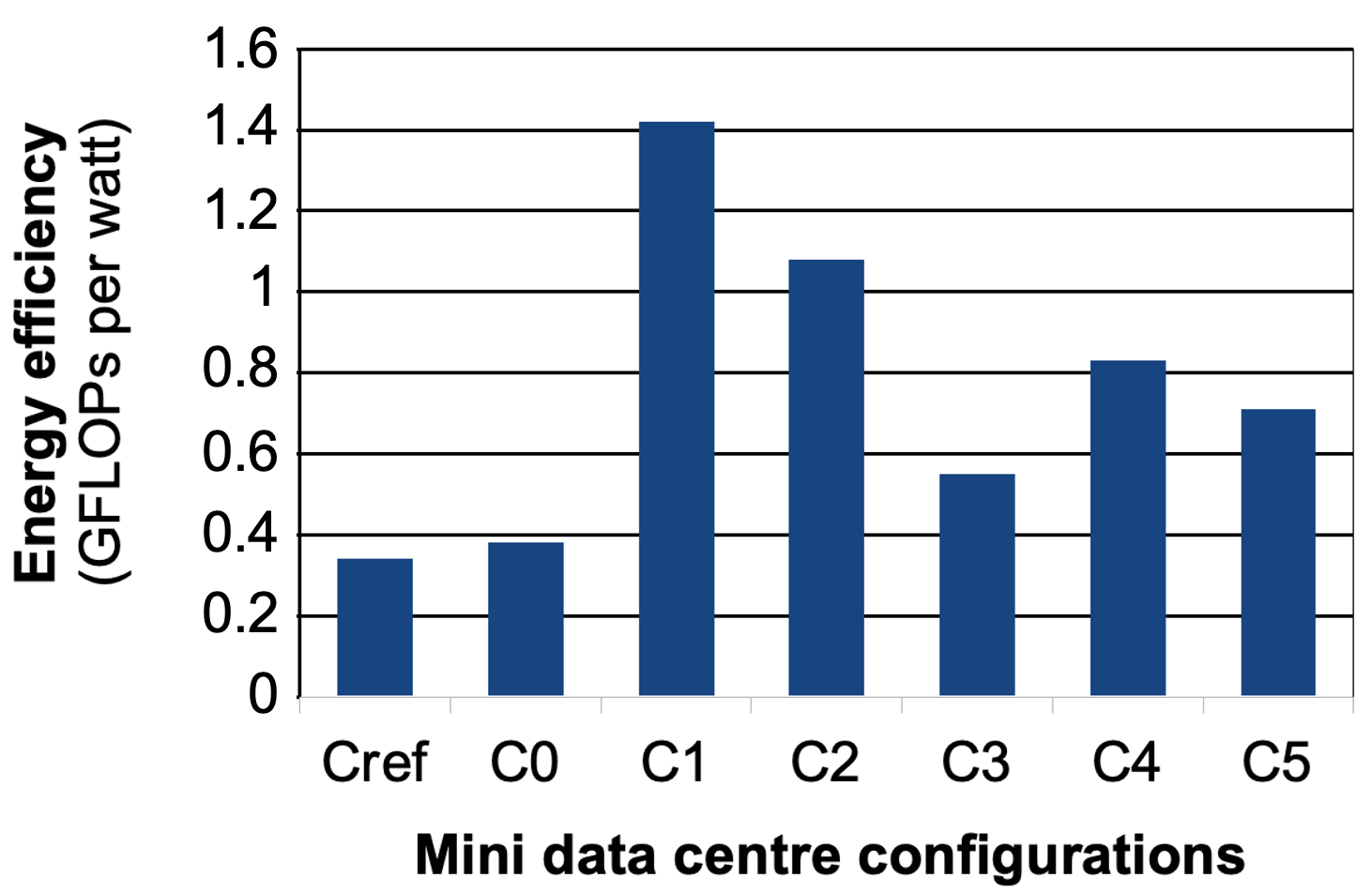}
         \caption{Energy efficiency}
         \label{fig:eeff}
     \end{subfigure}
     \hfill
     \begin{subfigure}[htbp]{0.45\textwidth}
         \centering
            \includegraphics[height=4.3cm]{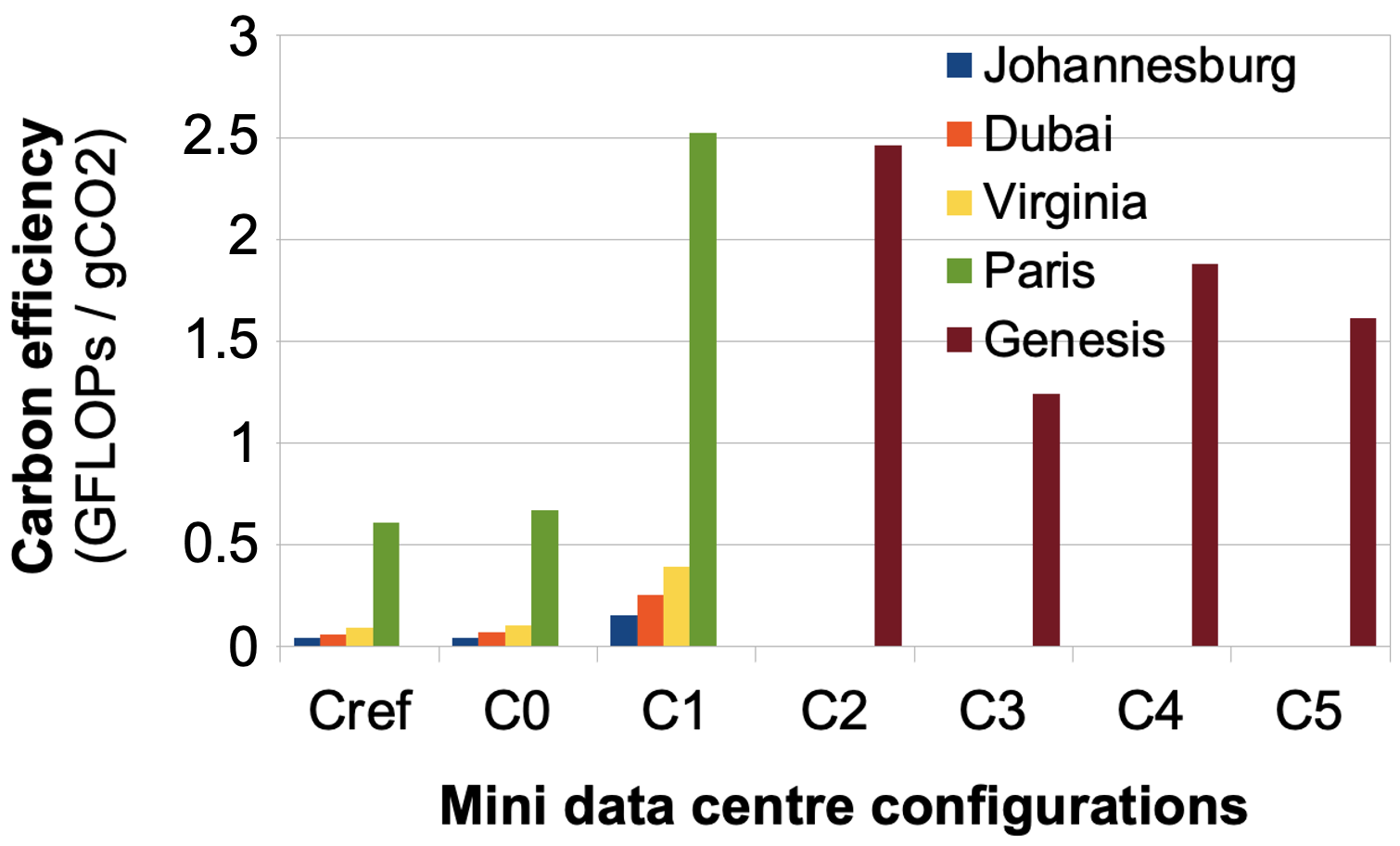}
         \caption{Carbon efficiency}
         \label{fig:ceff}
     \end{subfigure}
     \begin{subfigure}[htbp]{0.45\textwidth}
         \centering
            \includegraphics[height=4.1cm]{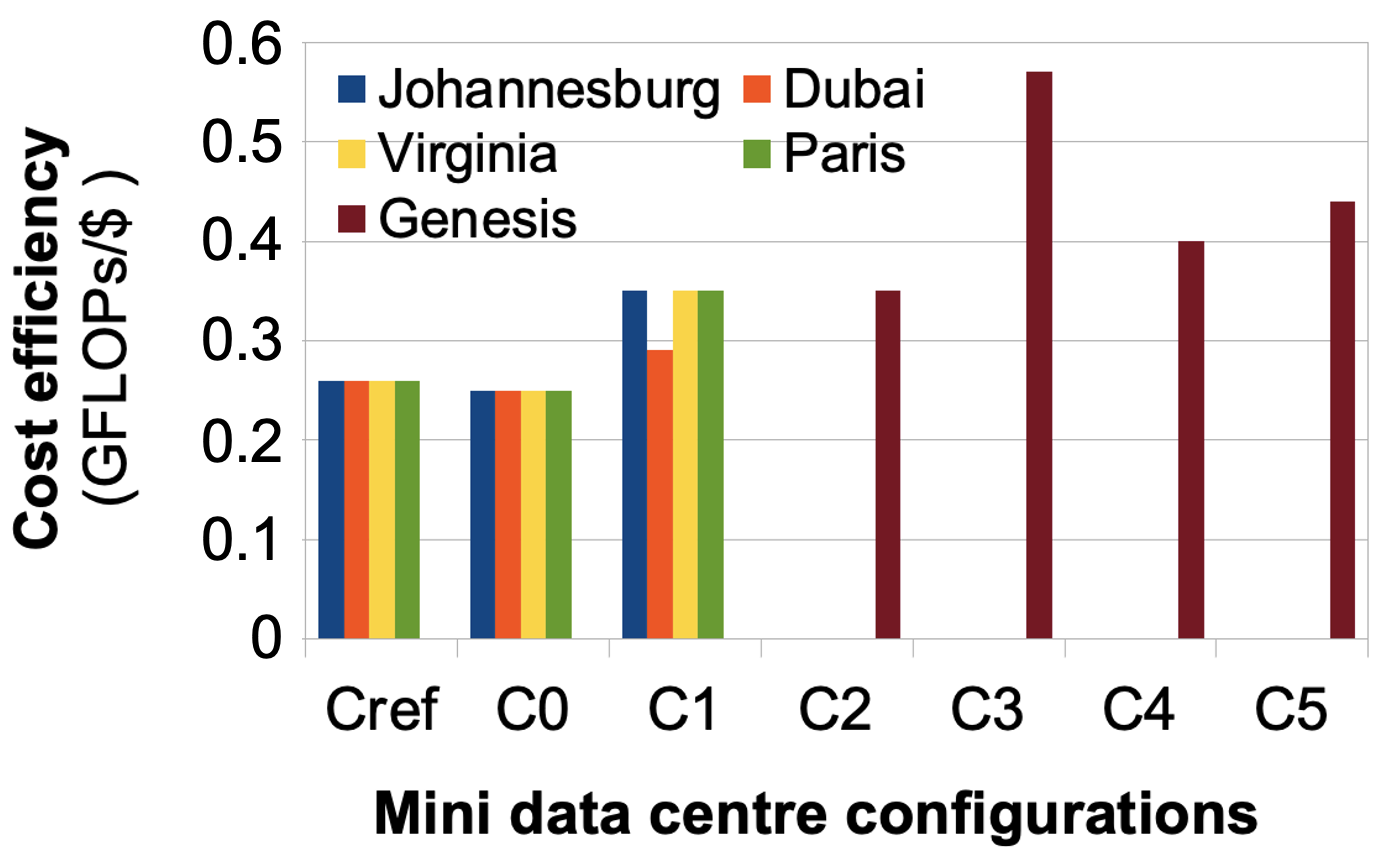}
         \caption{Cost efficiency}
         \label{fig:coeff}
     \end{subfigure}
        \caption{Three design efficiency metrics comparing classical server renewal w.r.t. Genesis approach}
        \label{fig:efficiency_metrics}
\end{figure*}

In the previous section, we evaluated Genesis technology configurations. This illustrated the opportunity to combine new and older generation servers in the same mini data centre for sustainability. On the other hand, by utilizing the green energy available in the batteries, a significant amount of computation can be supported by the system. 
Now, we compare the different configurations explored previously from the perspective of three strategic metrics (see Figure  \ref{fig:efficiency_metrics}): energy efficiency, i.e., computing capacity per power unit, carbon efficiency, i.e. compu\-ting capacity per $CO_2$ unit, and execution cost efficiency,  i.e. computing capacity per dollar.

\quad Figure \ref{fig:eeff} illustrates the energy efficiency of the different configurations. Configuration $C_1$ has the highest energy efficiency among others. However, Configuration $C_2$ also has high energy efficiency, and, most importantly, the energy used in this configuration is green energy. This is not the case with Configuration $C_1$.

\quad To assess the carbon efficiency of our study, we will consider various deployment scenarios of mini data centres in Johannesburg, Dubai, Virginia, and Paris. Carbon emissions for energy production vary from one city to another, so we will use data from \cite{silvavasconcelos:hal-04032094} research that quantifies carbon emissions by city. Solar energy sources emit 44 g$CO_2$/kWh according to \cite{CO2}. Looking at the carbon efficiency of the different configurations, $C_1$ provides the highest results when operating in Paris (see Figure \ref{fig:ceff}). However, it is only slightly better than $C_2$ in energy efficiency. Note that except $C_1$ in Paris, Genesis-based configurations outperform conventional data centre designs. Note that the carbon efficiency evaluation reported in Figure \ref{fig:ceff} does not take into account the additional carbon footprint associated with the server renewal strategies in $C_{ref}$, $C_0$, $C_1$ and $C_2$. Taking this missing carbon footprint into account can make $C_3$, $C_4$ and $C_5$ better.

\quad Figure \ref{fig:coeff} compares the cost of the considered configurations. This estimation comprises the cost of deploying servers according to Table \ref{tab:perform} and the cost of energy usage during execution. In general, Genesis configurations are more cost-effective than other configurations. This can be partially attributed to the re-purposing of outdated servers, which results in cost savings by acquiring new equipment. It must be noted that, for our efficiency analysis, we assumed that solar irradiation conditions are sufficient to power Genesis.

\quad Overall, we observe that the Genesis approach is a suitable candidate for dealing with server lifetime extension in mini data centres while mitigating the associated energy efficiency penalty that may result from the server aging effect. Most importantly, Genesis makes mini data centers more sustainable by providing better carbon efficiency.

\section{Concluding remarks and perspectives}
This work aimed to show how Genesis can reduce electronic waste emitted into the environment by data centres. During this analysis, we illustrated how data centres can positively evolve by integrating new servers and extending the lifetime of older servers in the system. Genesis' solar energy has less carbon impact on the environment than a large number of mini data centres that often rely on an energy mix inherent to the local utility grid, partly relying on fossil fuels such as coal. From a financial perspective, Genesis-oriented mini data centres can be configured to save money. We used analytical reasoning to show the potential benefits of Genesis over conventional mini data center designs w.r.t. server renewal. Overall, thanks to all these interesting features, Genesis appears as an excellent candidate for mini data center sustainability.

\quad Further perspectives to this work-in-progress is to refine the considered analytical modeling by relax\-ing some strong assumptions and by integrating more realistic operational conditions, e.g., server execu\-tion costs beyond the processor components, a higher diversity of processor power dissipation levels accord\-ing to executed load, analysis under various solar irradiation conditions, etc.  

\begin{small}
\bibliography{refs}
\end{small}
\appendix
\section{Appendix}
%\is{
To assess the carbon efficiency in our study, we considered various deployment scenarios of the mini data center in four cities: Johannesburg, Dubai, Virginia, and Paris. Carbon emissions for energy production vary from one city to another. So, we used data from \cite{silvavasconcelos:hal-04032094}, which quantifies carbon emissions according to city (see Table \ref{tab:emission}). According to the authors \cite{CO2}, solar energy sources emit 44 g$CO_2$/kWh. Regarding the electricity price, we used the statistics given by \cite{stat:2022} in 2022 per country worldwide, as indicated in Table \ref{tab:emission}.
%}

\begin{table}[h]
\footnotesize
   \centering
       \begin{tabular}{|p{3.5cm}||p{2cm}|p{2cm}|p{2cm}|p{2cm}|}
      \hline
       Location & Johannesburg & Dubai & Virginia & Paris \\
    \hline
    Emissions ($gCO_2$/kWh) & 900.6 & 530.0 & 342.8 &  52.6 \\
    \hline
     \hline
    Price (\$/kWh) & 0.15 & 0.8 & 0.18 & 0.21 \\
    \hline
\end{tabular} 
   \caption{$CO_2$ emission and energy cost numbers in the cities considered in our study}
   \label{tab:emission}  
\end{table}

\end{document}